\documentclass[12pt]{article}
\parskip=\medskipamount

\usepackage{tensor}
\usepackage[spanish,english]{babel}
\usepackage{physics}
\usepackage{slashed}
\usepackage{amsmath,amssymb,calc, amsthm,bbm, epsfig,psfrag, mathtools}
\usepackage[normalem]{ulem} 
\usepackage{jheppub}
\makeatletter
\def\@fpheader{\relax}
\makeatother

\usepackage{latexsym,bm,amsfonts}
\usepackage{graphicx, enumerate}
 \usepackage[all,cmtip]{xy}
\usepackage[numbers,sort&compress]{natbib}
\usepackage[dvipsnames]{xcolor}
\usepackage{mathrsfs}
\usepackage{tikz,tikz-cd}

\allowdisplaybreaks

\newcommand{\Comment}[1]{{}}
\definecolor{darkblue}{rgb}{0.15,0.35,0.55}
\definecolor{reddish}{rgb}{0.65, 0.2, 0.2}
\usepackage[linktocpage=true]{hyperref}
\hypersetup{
colorlinks=true,
citecolor=darkblue,
linkcolor=reddish,
urlcolor=darkblue,
pdfauthor={},
pdftitle={},
pdfsubject={}
}
\def\be{\begin{equation}}
\def\ee{\end{equation}}
\def\bea{\begin{eqnarray}}
\def\eea{\end{eqnarray}}

\title{ \boldmath Waves and strings in an interacting conformal chiral 2-form theory in six dimensions}

\author[a,b]{Nihat Sadik Deger,}
\author[c]{\'Angel J. Murcia,}
\author[c.d]{Dmitri Sorokin \vspace{0.15cm}}

\affiliation[a]{Department of Mathematics, Bogazici University,
Bebek, 34342, Istanbul, Turkey.
 \vspace{0.15cm}}
 \affiliation[b]{Feza Gursey Center for Physics and Mathematics, Bogazici University, Kandilli, 34684, Istanbul, Turkey.
 \vspace{0.15cm}}
\affiliation[c]{INFN, Sezione di Padova, Via Francesco Marzolo 8, 35131 Padova, Italy. \vspace{0.15cm}}
\affiliation[d]{Dipartimento di Fisica e Astronomia ``Galileo Galilei",  Universit\`a degli Studi di Padova,
Via Francesco Marzolo 8, 35131 Padova, Italy. \vspace{0.15cm}}

\emailAdd{sadik.deger@bogazici.edu.tr}
\emailAdd{angel.murcia@pd.infn.it}
\emailAdd{dmitri.sorokin@pd.infn.it $\,  \break $}

\date{\today}
\abstract{
In six dimensions there exists a unique one-parameter family of non-linear conformal electrodynamics for a chiral 2-form gauge field which includes (in a free-field limit) the linear chiral 2-form theory and is related, by dimensional reduction, to the four-dimensional ModMax electrodynamics.
In this work, we present the first exact solutions of this theory in presence of gravity. In particular, we will consider plane-wave, Robinson-Trautman, and self-dual dyonic string configurations, generalizing well-known solutions of this type in six-dimensional supergravities with linear chiral 2-form multiplets.
 }

\numberwithin{equation}{section}
\begin{document}
\maketitle
\flushbottom


\allowdisplaybreaks
\section{Introduction}
It was found in \cite{Bandos:2020jsw}   that there exists a unique non-linear extension of four-dimensional Maxwell's theory (dubbed ModMax) that preserves all symmetries of Maxwell's electrodynamics including electric-magnetic duality and conformal invariance. In view of its uniqueness, ModMax has attracted a great deal of attention and has been studied from different perspectives, in particular, in the context of black hole physics, Taub-NUT and pp-wave solutions (see e.g.  \cite{Flores-Alfonso:2020euz,BallonBordo:2020jtw,Flores-Alfonso:2020nnd,Amirabi:2020mzv,Bokulic:2021dtz, Kruglov:2022qag,Ali:2022yys,Kubiznak:2022vft,Barrientos:2022bzm,Bokulic:2022cyk,Amirabi:2022hsv,Pantig:2022gih,Bakhtiarizadeh:2023mhk,Guzman-Herrera:2023zsv,Sharif:2023psx,Tahamtan:2023tci,Ayon-Beato:2024vph,Shahzad:2024ljt}), and as a characteristic example that triggered the development of new marginal (so called root-$T\bar T$) deformations of field theories in various dimensions \cite{Rodriguez:2021tcz,Babaei-Aghbolagh:2022uij,Ferko:2022iru,Ferko:2022cix,Conti:2022egv,Borsato:2022tmu,Tempo:2022ndz,Ferko:2023ruw,Garcia:2022wad,Ferko:2023ozb,Ferko:2023iha,Ferko:2024zth,Ebert:2024zwv,Morone:2024ffm,Babaei-Aghbolagh:2024hti,Tsolakidis:2024wut,Hadasz:2024pew}. It has Born-Infeld-like deformations \cite{Bandos:2020hgy,Kruglov:2021bhs,Nastase:2022fzx}, as well as supersymmetric \cite{Kuzenko:2021cvx,Bandos:2021rqy}, higher-spin  \cite{Kuzenko:2021qcx}, sigma-model \cite{Kuzenko:2023ysh} and higher-derivative \cite{Kuzenko:2024zra} generalizations. Furthermore, ModMax is a causal theory \cite{Bandos:2020jsw,Russo:2024kto,Russo:2024llm} and its non-linear equations of motion  have exact solutions including plane waves \cite{Bandos:2020jsw}, generic null electromagnetic fields \cite{Dassy:2021ulu} and Lienard-Wiechert fields induced by moving dyonic particles  \cite{Lechner:2022qhb}. Birefringence and Compton effects have also been studied in this theory \cite{Flores-Alfonso:2020euz,Bandos:2020jsw,Lechner:2022qhb,Shi:2024nmx}.

In six space-time dimensions there exists a close relative of ModMax: a unique non-linear conformal chiral 2-form theory \cite{Bandos:2020hgy} which reduces to the former upon a dimensional reduction. It contains a dimensionless parameter $\gamma$ such that for $\gamma=0$ the theory reduces to the free chiral 2-form electrodynamics. This six-dimensional ModMax counterpart and its Born-Infeld-like generalization \cite{Bandos:2020hgy} have been subject so far to a much less detailed study, one of which was (marginally) in the context of the trirefringence phenomenon \cite{Bandos:2023yat} and another one in the context of $T\bar T$-like deformations of six-dimensional chiral 2-form theories \cite{Ferko:2024zth}. In this paper we initiate the systematic study of exact solutions of this theory, which we will call `$6D$ ModMax' for short. In particular, we will concentrate on wave-like and black-string solutions of this theory coupled to gravity, inspired by well-known solutions of this kind in $D=6$ supergravities containing linear chiral 2-form multiplets.

Among wave configurations in the $6D$ Einstein-ModMax theory, we shall explore a generalization of pp-wave solutions \cite{Hely:1959,Peres} studied in the context of $D=6$ supergravity in \cite{Meessen:2001vx,Gutowski:2003rg}, Siklos space-times \cite{Siklos,Gibbons:1986uv} (gravitational-wave spaces with planar wavefronts in AdS) and Robinson-Trautman solutions associated with the presence of gravitational radiation \cite{Robinson:1962zz,Robinson:1983ws}, considered previously in D-dimensional Einstein gravity minimally coupled to linear p-form gauge theories  \cite{Ortaggio:2014gma,Ortaggio:2016zuk}.  As for Siklos space-times, they have been generalized to arbitrary dimensions  in the absence of matter fields \cite{Gibbons:1986uv}, and here we will be interested in the addition to the Einstein equations of a 2-form gauge field satisfying the non-linear self-duality condition of $6D$ ModMax. To the best of our knowledge these solutions have not appeared in the literature yet even in the linear case of $6D$ supergravity chiral 2-forms, probably because they do not preserve supersymmetry.

Regarding black-string configurations, a supersymmetric self-dual black string solution was first found in un-gauged $D=6$ supergravity in \cite{Duff:1993ye}\footnote{A similar self-dual string soliton solution exists in the non-linear chiral 2-form theory on the worldvolume of the $M5$-brane \cite{Howe:1997ue}.}
and its generalization with independent electric and magnetic charges appeared in
\cite{Duff:1995yh} (see also \cite{Duff:1996cf,Duff:1996hp,Duff:1998cr}). Interestingly enough, this solution was shown to be an uplift of a static, uncharged string solution of a $D=3$ gauged supergravity \cite{Deger:2019jtl}. We will focus on the study of a static black-string configuration and its thermodynamics in the $6D$ Einstein-ModMax theory. We find that a main effect of the ModMax non-linearity is a change in the relation between the AdM mass and charge of the extremal self-dual string which involves the ModMax parameter $\gamma$ similar to what happened for $4D$ Reissner-Nordstr\"om ModMax black holes \cite{Flores-Alfonso:2020euz,BallonBordo:2020jtw,Flores-Alfonso:2020nnd}. This relation also resembles the effect of a constant dilaton in a self-dual string solution of a $6D$ supergravity discussed in \cite{Duff:1993ye,Duff:1996cf}. As for rotating dyonic string solutions, their existence and supersymmetric properties have been studied in various works, see e.g. \cite{Cvetic:1998xh,Gueven:2003uw,Randjbar-Daemi:2004bjl,Jong:2006za}. In this paper we focus on a near-horizon geometry of a particular subclass of rotating black strings considered in \cite{Chow:2019win,Pang:2019qwq}, for which we find a ModMax generalization.

The paper is organized as follows. In Section \ref{sec:2} we  start with a brief description of the 6D ModMax theory in a curved space-time, presenting its Lagrangian, equations of motion and the energy-momentum tensor. In Section \ref{sec:couplingstring} we present a procedure of how to consistently couple $6D$ ModMax to a dynamical self-dual string, as well as the equations of motion describing the Einstein-ModMax system.  In Section \ref{NPP} generic aspects of null-field configurations in $6D$ ModMax theory are considered. These results are then used in Section \ref{sec:planewaves}, in which novel wave-like solutions of the Einstein-ModMax system are presented. In Section \ref{sec:static} we obtain a static black string solution in the Einstein-ModMax theory and show that it satisfies the first law of  thermodynamics. We also obtain an extremal self-dual string solution with waves propagating along the string worldsheet. In Section \ref{sec:rotbh} we present another solution of the Einstein-ModMax system which we interpret as a \emph{near-horizon} geometry of a rotating extremal black string. We conclude by discussing open problems and possible future directions.

In the Appendix we give details on the relation between two Lagrangian formulations of $N=(1,0)$, $D=6$ supergravity coupled to a single tensor multiplet, the original one \cite{Marcus:1982yu} involving a non-chiral 2-form gauge field and the PST-formulation \cite{Dall'Agata:1997db}, which uses one chiral and one anti-chiral 2-form field. To our knowledge, the comparison of these two formulations have not been discussed in the literature yet.

{\bf Notation and conventions.} We use lower case Greek letters as six-dimensional space-time indices $\mu,\nu,\ldots=(0,1,\ldots,5)$, and lower case Latin letters as $5$-dimensional spatial indices  $i,j,\ldots=(1,\ldots,5)$. The signature of the Minkowski metric is chosen to be ``almost plus". The (anti-)symmetrization of indices is performed with weight one, i.e. $X_{[\mu_1\mu_2\ldots\mu_p]}=\frac 1{p!}(X_{\mu_1\mu_2\ldots \mu_p}-X_{\mu_2\mu_1\ldots \mu_p}+\cdots)$ and $X_{(\mu_1\mu_2\ldots\mu_p)}=\frac 1{p!}(X_{\mu_1\mu_2\ldots \mu_p}+X_{\mu_2\mu_1\ldots \mu_p}+\cdots)$. The $6D$ totally antisymmetric tensor density $\varepsilon_{\mu_1\ldots\mu_6}$ is such that
$
\varepsilon_{012345}=-\varepsilon^{012345}=1.
$

\section{Preliminaries: PST formulation of the $6D$ ModMax theory}
\label{sec:2}
The PST Lagrangian density for a generic self-interacting 2-form field $A_{\mu\nu}$ (of canonical dimension of mass squared) in a curved $6D$ space-time with a metric $g_{\mu\nu}$ has the following form \cite{Pasti:1997gx,Buratti:2019guq,Bandos:2020hgy}
\be\label{PST}
\mathcal L_{PST}=\sqrt{-g}\left(\frac 14 B_{\mu\nu}F^{\mu\nu\rho}v_\rho-\mathcal H(s,p)\right)\,,
\ee
where $g=\det g_{\mu\nu}$, $F_{\mu\nu\rho}=3\partial_{[\mu}A_{\nu\rho]}$
and
\be\label{Bp}
B^{\mu\nu}=\frac 1{2\sqrt{-g}}\varepsilon^{\mu\nu\rho\lambda\sigma\kappa}\partial_{\lambda}A_{\sigma\kappa}v_\rho
=F^{*\mu\nu\rho}v_\rho.
\ee
Observe that $B^{\mu \nu} v_{\mu}=0$. The vector $v_\mu$ is a normalized gradient of an auxiliary scalar field $a(x)$, such that $v_\mu$ is a unit time-like vector which, in the almost plus metric convention, is
\be\label{vmu}
v_{\mu}v^\mu=-1, \qquad v_\mu=\frac{\partial_\mu a}{\sqrt{-\partial_\nu a\partial^\nu a}}\,.
\ee
The term $\mathcal H(B_{\mu\nu})=\mathcal H(s,p)$ is a function of  two Lorentz (or general-coordinate) invariants $s$ and $p$, which are the only two independent invariant combinations of $B_{\mu\nu}$ (obeying $B_{\mu\nu}v^\nu=0$) at any order of $B$, defined as follows
\bea
\label{ps}
s&=&\frac 14 B_{\mu\nu}B^{\mu\nu}\,,\nonumber\\
p^{\mu} &=&-\frac 1{8\sqrt{-g}}\varepsilon^{\mu\nu\rho\lambda\sigma\kappa}B_{\rho\lambda}B_{\sigma\kappa}v_{\nu}\,, \qquad p=\sqrt{p_\mu p^\mu},
\eea
For the $6D$ ModMax theory
\be\label{MMPST}
\mathcal H_{MM}=\cosh(\gamma)\,s-\sinh(\gamma)\,\sqrt{s^2-p^2}\,,
\ee
from which the free chiral-two form case is obtained by setting $\gamma=0$. Let us also define
\be\label{E}
F_{\mu\nu\rho}v^{\rho}=:E_{\mu\nu}\,.
\ee
From their definitions, we may readily note that $E_{\mu\nu}$ and $B_{\mu\nu}$ are analogous of the electric and magnetic fields of $4D$ electrodynamics.

In addition to the conventional gauge invariance $\delta A_{\mu\nu}=2\partial_{[\mu}\lambda_{\nu]}$, the local symmetries of the PST action \eqref{PST} are
\be\label{PST1}
\delta A_{\mu\nu}=2\partial_{[\mu}a\,\Phi_{\nu]}\,, \qquad \delta a=0,
\ee
and
\be\label{PST2}
\delta a =\varphi(x), \qquad \delta {A_{\mu\nu}}=-\frac{\varphi}{\sqrt{-\partial_\mu a\partial^\mu a}}(E_{\mu\nu}- H_{\mu\nu})\,.
\ee
where $\Phi_\nu(x)$ and $\varphi(x)$ are local symmetry parameters,
\bea\label{Hmn}
H_{\mu\nu}&=&2\frac{\partial{\mathcal H}}{\partial B^{\mu\nu}}=({\cal H}_s + 2s p^{-1} {\cal H}_p) B_{\mu\nu} + p^{-1}{\cal H}_p (B^3)_{\mu\nu}\, ,
\eea
and $(B^3)_{\mu\nu}=B_{\mu \lambda} B^{\lambda \rho} B_{\rho \nu}$. In \eqref{Hmn} and below ${\cal H}_s =\partial_s \mathcal H$ and ${\cal H}_p =\partial_p \mathcal H$.

The action \eqref{PST} is invariant under the local transformations \eqref{PST2} provided the function $\mathcal H(s,p)$ satisfies the condition
\be\label{Li}
{\cal H}^2_s + \frac{2s}p{\cal H}_s{\cal H}_p + {\cal H}^2_p  =1 \, .
\ee

The local shifts \eqref{PST2} of the scalar $a(x)$ can be used to impose
the gauge $\partial_\mu a=\delta^0_\mu$. \footnote{Note that because of the definition of $v_\mu$ in \eqref{vmu} the gauge $a=const$ is not admissible. As explained in Section 6.1 of \cite{Ferko:2024zth}, this restriction implies non-trivial conditions on the global causal structure of space-time, which must be globally hyperbolic \cite{Bernal:2004gm,bar_notes}. It is known that this space-time property is necessary for defining the Hamiltonian (ADM) formulation of General Relativity. In the PST formulation it ensures the existence of the Hamiltonian formulation for chiral $p$-forms.} Then, if for simplicity we consider the flat space-time case, $B^{\mu\nu}$ defined in \eqref{Bp} reduces to $-B_{ij}$ and $E_{\mu\nu}$ defined in \eqref{E} reduces to $-E_{ij}$ of the Hamiltonian formulation, and the PST Lagrangian density reduces to the first-order time derivative Lagrangian density of the Hamiltonian formulation of the chiral-form theory\footnote{Upon a dimensional reduction to $D=4$ this Lagrangian density and the corresponding Hamiltonian density reduce to the Hamiltonian formulation of a $4D$ non-linear electrodynamics, as explained in \cite{Bandos:2020hgy}. }
\be\label{LH}
\mathcal L_H=\frac 14 B_{ij}\partial_0 A^{ij}- \mathcal H(B_{ij})\,.
\ee
In the generic non-linear chiral 2-form theory
the PST equations of motion of $A_{\mu\nu}$ and $a(x)$ are, respectively,
\be\label{Aem}
\varepsilon^{\mu\nu\rho\sigma\kappa\lambda}\partial_\rho \Big(v_{\sigma}(E_{\kappa\lambda}-H_{\kappa\lambda})\Big)=0
\ee
and
\be\label{aem}
\varepsilon^{\mu\nu\rho\sigma\kappa\lambda}(E_{\mu\nu}-H_{\mu\nu})\partial_\rho \Big(v_{\sigma}(E_{\kappa\lambda}-H_{\kappa\lambda})\Big)=0\,.
\ee
Note that the $a(x)$-field equation  is not independent. It holds whenever the \hbox{$A_{\mu\nu}$-field} equation \eqref{Aem} is satisfied. This reflects the fact that $a(x)$ is a non-dynamical pure gauge field.

It can then be shown, with the use of the local symmetry \eqref{PST1}, that eq. \eqref{Aem} produces, upon integration, the non-linear self-duality condition
\be\label{PSTsd}
E_{\mu\nu}=H_{\mu\nu}.
\ee
Note that, on the mass-shell \eqref{PSTsd} the PST Lagrangian density \eqref{PST} is\footnote{To derive \eqref{PSTonshell} one should use the expression \eqref{Hmn} and the identity $\frac{1}{4} (B^3)_{\mu \nu} B^{\mu \nu}=p^2-2s^2$.}
\be\label{PSTonshell}
\mathcal L_{PST}^{on-shell}=\frac 14 B^{\mu\nu}H_{\mu\nu}-\mathcal H=(s\mathcal H_s+p\mathcal H_p)-\mathcal H(s,p).
\ee
It vanishes in the free theory and in the ModMax case due to the conformal invariance, which implies that $\mathcal H$ is a homogeneous function of $s$ and $p$ of order one, i.e.
\be\label{confi}
\mathcal H_{conf}(s,p)=s\mathcal H_s+p\mathcal H_p\,.
\ee
A generic chiral 2-form energy-momentum tensor has the following form\footnote{It was first derived in \cite{Bandos:1997gm} for the description of the M5-brane equations of motion.}
\be\label{EMPST}
T_{\mu\nu}=-\frac 2{\sqrt{-g}}\frac{\partial \mathcal L}{\partial g_{\mu\nu}}=-g_{\mu\nu}\left(\mathcal H+\frac 12\tr HB\right )-\frac 12 v_\mu v_\nu \tr HB +(HB)_{\mu\nu}- 2 v_{(\mu}p_{\nu)}.
\ee
 On the mass-shell \eqref{PSTsd} it reduces to \cite{Sorokin:1998xx}
\be\label{EMPSTOS2}
T_{\mu\nu}|_{on-shell}=g_{\mu\nu}\left(\frac 14\,H_{\rho\lambda}B^{\rho\lambda}-\mathcal H\right) +\frac 14F_{\rho\lambda(\mu}F_{\nu)}^{*\,\,\rho\lambda}=g_{\mu\nu} \,\mathcal L|_{on-shell} +\frac 14 F_{\rho\lambda(\mu}F_{\nu)}^{*\,\,\rho\lambda}\,.
\ee
As we have already seen, for ModMax $\mathcal L|_{on-shell}=0$, so its on-shell energy momentum tensor is
\be\label{OSETMM}
T^{MM}_{\mu\nu}|_{on-shell}=\frac 14 F_{\rho\lambda(\mu}F_{\nu)}^{*\,\,\rho\lambda}\,.
\ee
This was shown \cite{Ferko:2024zth} to be equivalent to
\be\label{MMEMPSTOS}
T^{MM}_{\mu\nu}|_{on-shell}=\frac 14F_{\rho\lambda(\mu}F_{\nu)}^{*\,\,\rho\lambda}= \frac 1{4\cosh\,\gamma}\left(F_{\mu\rho\lambda}F_{\nu}{}^{\rho\lambda}-\frac 16\eta_{\mu\nu}F^2\right).
\ee
 For the free chiral 2-form theory (corresponding to $\gamma=0$) the 3-form field strength is self-dual
\be\label{sd}
F_{\mu\nu\rho}=F^*_{\mu\nu\rho},
\ee
and hence on the mass-shell the free (self-dual) chiral-field energy momentum tensor is
\be\label{T0}
T^0_{\mu\nu}=\frac 14 F_{\mu\lambda\rho}F_{\nu}{}^{\lambda\rho}\,.
\ee
We will use the obtained manifestly covariant form of the on-shell ModMax energy momentum tensor \eqref{MMEMPSTOS} for the study of solutions of this theory coupled to gravity.

Finally, let us present the equations of motion for the ModMax chiral 2-form theory. They can be written in the following forms which are almost (but not completely) equivalent, as we explain below eq. \eqref{tilde T}. One is obtained from the non-linear self-duality condition \eqref{PSTsd} for $\mathcal H$ given in \eqref{MMPST} and involves the auxiliary vector $v_\mu$
\be\label{MMPSTsd}
E_{\mu\nu}=\frac 1{\sqrt{s^2-p^2}}\left(\cosh\gamma{\sqrt{s^2-p^2}}+s\,\sinh\gamma\right) B_{\mu\nu} +\frac{ \sinh\gamma}{\sqrt{s^2-p^2}} (B^3)_{\mu\nu}\,.
\ee
Another one is the manifestly covariant $v^\mu$-independent equation which is obtained upon some algebraic manipulations from a general non-linear self-duality equation considered in \cite{Avetisyan:2022zza} and \cite{Ferko:2024zth}:
\be\label{eomMk'''}
F^{*\,\mu\nu\rho}=F^{\nu\rho\lambda}\left(\cosh\gamma \,\delta_\lambda^\mu-\frac {24\,{\sinh^2\gamma}\,}{F^2}\,T_\lambda{}^{\mu}\right):=G^{\mu\nu\rho}\,,
\ee
or, alternatively,
\be\label{eomMk}
F^{\,\mu\nu\rho}=F^{*\,\nu\rho\lambda}\left(\cosh\gamma \,\delta_\lambda^\mu-\frac {24\,{\sinh^2\gamma}\,}{(F^*)^2}\,\tilde T_\lambda{}^{\mu}\right):=G^{*\mu\nu\rho}\,,
\ee
where
\be\label{tilde T}
\tilde T_{\mu\nu}=\frac 1{4\cosh\,\gamma}\left(F^*_{\mu\rho\lambda}F^*_{\nu}{}^{\rho\lambda}-\frac 16\,g_{\mu\nu}(F^*)^2\right)\equiv T_{\mu\nu}\,.
\ee
With the use of the identity
\cite{Bandos:1997gm}
\be\label{id}
F^{\mu\nu\rho}=-\frac 1{2\sqrt{-g}}\varepsilon^{\mu\nu\rho\lambda\sigma\delta}v_\lambda B_{\sigma\delta}-3v^{[\mu}E^{\nu\rho]},
\ee
from \eqref{MMPSTsd} one finds that on the mass-shell
\bea\label{FF=MM}
F_{\mu\nu\rho}F^{\mu\nu\rho}&=&3(B_{\mu\nu}B^{\mu\nu}-E_{\mu\nu}E^{\mu\nu})=
3 (4s-H_{\mu\nu}H^{\mu\nu})=24\frac{s^2-p^2}p\mathcal H_s\mathcal H_p\nonumber\\
&=& 24\sinh\gamma\,\sqrt{s^2-p^2}\,\mathcal H_s
= 24(s-\cosh\gamma\,\mathcal H_{MM})\,.
\eea
Projecting eqs. \eqref{eomMk'''} and \eqref{eomMk} along $v_\mu$, upon some algebraic manipulations one gets eq. \eqref{MMPSTsd}. However, it should be pointed out that the equations \eqref{eomMk'''} and \eqref{eomMk} include the cases of the both signs $\mp$ of $\sinh\gamma$ in the definition \eqref{MMPST} of $\mathcal H_{MM}(s,p)$ and correspondingly in \eqref{MMPSTsd}. These two choices are related by the replacement of $\gamma$ with $-\gamma$ under which \eqref{eomMk'''} and \eqref{eomMk} are invariant. Another related subtlety is that the equations \eqref{eomMk'''} and \eqref{eomMk} are not suitable for the description of null field configurations with $F_{\mu\nu\rho}F^{\mu\nu\rho}=0$, such as plane waves. In this case eq. \eqref{MMPSTsd} is indispensable. As follows from the relation \eqref{FF=MM}, for the choice of the sign of $\sinh \gamma$ as it appears in \eqref{MMPST} and \eqref{MMPSTsd}, the null field solutions exist only for $\gamma\geq 0$, while for the opposite choice of the sign of $\sinh\gamma$ the null field solutions exist if $\gamma\leq 0$. To fix this sign ambiguity, in what follows we will only consider the solutions of the ModMax theory described by the function \eqref{MMPST} with $\gamma\geq 0$. This choice is in accordance with the absence of superluminal propagation of small wave disturbances on constant uniform field backgrounds in $4D$  \cite{Bandos:2020jsw} and $6D$ \cite{Bandos:2023yat} ModMax.

\section{Coupling $6D$ ModMax to a dynamical self-dual string}
\label{sec:couplingstring}
In \cite{Deser:1997mz,Medina:1997fn,Bandos:1997gd,Lechner:2000eg} it was shown that linear chiral $p$-forms can couple to dynamical self-dual $p$-branes (carrying equal electric and magnetic charges) in the same way as  dyons couple to electromagnetic fields in $4D$ electrodynamics or to fields related by a twisted self-duality (see \cite{Malcha:2024bts} for a recent discussion). This coupling is straightforwardly generalized to the case of interacting chiral $p$-form theories and in particular to $6d$ ModMax in which case we are dealing with a self-dual string. The procedure of adding to the chiral 2-form action a string source is carried out as follows. We assume that  the coupling of the chiral 2-form field action \eqref{PST} to a string includes the Nambu-Goto string action and a minimal coupling of $A_{\mu\nu}$ to a string electric current
\be\label{minA}
S=S_{PST}-\frac 12\int \mathrm{d}^6x(\sqrt{-g}\, A_{\mu\nu}j^{\mu\nu})-T\int_{\mathcal W_2}\mathrm{d}^2\xi\sqrt{-\det\big(\partial_\alpha y^\mu\partial_\beta y^\nu g_{\mu\nu}(y(\xi))\big)},
\ee
where
\be\label{sj}
j^{\mu\nu}=\frac{\mathcal Q}{\sqrt{-g(x)} }\int_{\mathcal W_2} \mathrm{d}y^\mu \wedge \mathrm{d}y^\nu \delta^6 (x-y(\xi))\,,
\ee
is the string current, $\xi^\alpha=(\tau,\sigma)$ ($\alpha=0,1$) parameterize the string worldsheet $\mathcal W_2$, $y^\mu(\xi)$ are string embedding coordinates in $6D$, $\mathcal Q$ is the string charge and $T$ is its tension.

However, the addition of the minimal coupling term breaks the local symmetry \eqref{PST1}. To restore this symmetry one should modify the chiral form field strength $F_{3}=dA_2$ by adding to it a so-called Dirac membrane\footnote{The Dirac membrane is not physical and is the $6D$ counterpart of the $4D$ Dirac string introduced by Dirac for the description of monopoles. For details regarding the invisibility of Dirac branes and string/brane charge quantization we refer the reader to \cite{Deser:1997mz,Lechner:2000eg}.} whose boundary is the string, namely
\be\label{hatF}
\hat F_{\mu\nu\rho}=F_{\mu\nu\rho}-2C^*_{\mu\nu\rho}\,,
\ee
where
\be\label{C}
C^{\mu\nu\rho}=\frac{\mathcal Q}{\sqrt{-g(x)}}\int_{\mathcal M_3}\mathrm{d}z^\mu \wedge \mathrm{d}z^\nu\wedge \mathrm{d}z^\rho\delta^6(x-z(\omega))\,,
\ee
$\mathcal M_3$ being the three-dimensional worldvolume of the Dirac membrane whose boundary is the string worldsheet $\mathcal W_2$. $\mathcal M_3$ is parameterized by three coordinates $\omega^A=(\xi^\alpha,\omega^2)$ and $z^\mu(\omega)$ are embedding coordinates of $\mathcal M_3$ into six-dimensional space-time
such that $z^\mu(\xi^\alpha,0)=y^\mu(\xi)$.
The Dirac membrane is served to endow the string with the magnetic charge whose value coincides with its electric charge $\mathcal Q$, since
by construction, the following relation holds
\be\label{dC=j}
D_\mu C^{\mu\nu\rho}=-j^{\nu\rho},
\ee
where $D_\mu$ is the covariant derivative with respect to the metric-compatible connection in $D=6$.

The equivalence of the electric and magnetic charge is required for  the  action \eqref{PST} --- in which $E_{\mu\nu}$ and $B_{\mu\nu}$ are replaced with $\hat E_{\mu\nu}=\hat F_{\mu\nu\rho}v^\rho$ and $\hat B_{\mu\nu}=\hat F^*_{\mu\nu\rho}v^\rho$ --- to be invariant under the local symmetries \eqref{PST1} and \eqref{PST2}. Therefore, as described in the previous section,  the equations of motion of the chiral 2-form field derived from the action \eqref{minA} again reduce to non-linear self-duality conditions which have exactly the same form as in \eqref{MMPSTsd}-\eqref{eomMk} but with $F_{\mu\nu\rho}$ replaced by $\hat F_{\mu\nu\rho}$. Then, from \eqref{hatF}, \eqref{dC=j} and \eqref{eomMk'''} it follows that
\be\label{dhatG}
\frac 12 D_\mu \hat F^{*\mu\nu\rho}=\frac 12 D_\mu\hat G^{\mu\nu\rho}=j^{\nu\rho}\,.
\ee
We have thus coupled the ModMax theory to the self-dual string which carries the equal electric and magnetic charge $\mathcal Q$. The string electric current minimally couples to the chiral 2-form $A_{\mu\nu}$ and its magnetic current couples to the chiral field strength $F_{\mu\nu\rho}$ via the Dirac membrane.

We will now look for solutions in the $6D$ ModMax theory coupled to gravity, restricting ourselves to space-time regions away from the string sources.
To this end we will need the Einstein equations (with a cosmological constant $\Lambda$) sourced by the ModMax energy-momentum tensor
\be\label{ET}
R_{\mu\nu}-\frac 12 g_{\mu\nu}(R-2\Lambda)=8\pi G_N\, T_{\mu\nu}\,,
\ee
where $G_N$ is the Newton constant of the dimension of $(length)^4$. Since the ModMax energy-momentum tensor \eqref{MMEMPSTOS} is traceless,  we see that $R=3\Lambda$
and the equation \eqref{ET} takes the following explicit form
\be\label{EinsteinMM}
R_{\mu\nu}-\frac 12 g_{\mu\nu}\Lambda=2\pi G_N \,F_{\rho\lambda(\mu}F_{\nu)}^{*\,\,\rho\lambda}= \frac {2\pi G_N}{\cosh\,\gamma}\left(F_{\mu\rho\lambda}F_{\nu}{}^{\rho\lambda}-\frac 16\,g_{\mu\nu}F^2\right).
\ee

\section{Null fields and plane waves} \label{NPP}

Let us study whether the non-linear conformal chiral 2-form theory under consideration admits plane wave solutions. In flat space-time, these have the following  form
\be\label{pw}
F_{\mu\nu\rho}=f_{\mu\nu\rho}e^{ik_\mu x^\mu}+\bar f_{\mu\nu\rho} e^{-ik_\mu x^\mu},
\ee
where $f_{\mu\nu\rho}$ is a complex constant amplitude, $\bar f_{\mu\nu\rho}$ is its complex conjugate and $k_\mu$ is the wave vector.
In the linear case, self-duality \eqref{sd} of $F_{\mu\nu\rho}=3\partial_{[\mu} A_{\nu\rho]}$ implies that for the plane waves the following relations hold
\be\label{f=f^*}
f_{\mu\nu\rho}=f^*_{\mu\nu\rho}\,,
\ee
\be\label{F2=0}
F_{\mu\nu\rho}F^{\mu\nu\rho}=0,
\ee
\be\label{kf=0}
k^\mu f^*_{\mu\nu\rho}=0\,,
\ee
\be\label{k2=0}
k^\mu k_\mu=0\,.
\ee
We are interested in exploring whether there exist plane wave solutions satisfying the non-linear self-duality condition  \eqref{MMPSTsd} instead of \eqref{sd} (or \eqref{f=f^*}), while preserving all other properties \eqref{F2=0}-\eqref{k2=0}. The fields satisfying  \eqref{F2=0} are six-dimensional counterparts of the so-called null electromagnetic fields in four-dimensions whose $4D$ field strengths $F_{\hat\mu\hat\nu}$ ($\hat\mu,\hat\nu=0,1,2,3$) satisfy $F_{\hat\mu\hat\nu}F^{\hat\mu\hat\nu}=F_{\hat\mu\hat\nu}F^{*\hat\mu\hat\nu}=0$. The electromagnetic plane waves are examples of such fields.

Note that the form of the ModMax self-duality condition given in \eqref{eomMk'''}, having $F^2$ in the denominator, is not suitable for studying the null fields. So, for this purpose we will use the self-duality relation \eqref{MMPSTsd}.

The null-field condition \eqref{F2=0}, in view of \eqref{FF=MM}, implies the following relation between the invariants $s$ and $p$
\be\label{sprelation}
\sinh\gamma\, s- \cosh\gamma\,\sqrt{s^2-p^2}=0\,\quad \to \quad s^2=\cosh^2\gamma\,\, p^2 \quad \to \quad s=p\,\cosh\gamma \,.
\ee
Since $s$ and $p$ are positive definite the solution exists only for $\gamma\geq 0$ as in the $4D$ ModMax theory \cite{Bandos:2020jsw}. From the above equation and the form \eqref{MMPST} of the ModMax function $\mathcal H_{MM}$, we also see that $\mathcal H_{null}=p$ for null fields, which is the Hamiltonian density of the $6D$ counterpart \cite{Townsend:2019ils} of the $4D$ conformal Bialynicki-Birula electrodynamics \cite{BialynickiBirula:1992qj}.

Using the relation \eqref{sprelation} in \eqref{MMPSTsd} we get
\be\label{NFsd}
E_{\mu\nu}=2\cosh\gamma\,\left( B_{\mu\nu} +\frac{1}{2s} (B^3)_{\mu\nu}\right)\,.
\ee
Substituting this expression into the identity \eqref{id}
we find the expression for the field strength of a generic null field in this theory in terms of the components of $B_{\mu\nu}$ subject to the constraint \eqref{sprelation}. We should stress that equation \eqref{NFsd} is valid in any curved space-time background.

Let us now consider a plane wave solution \eqref{pw} in flat space-time. First, for simplicity, let us fix the PST gauge of the local symmetry \eqref{PST2} such that $v_\mu=-\delta_\mu^0$. Then, the non-zero components of $E_{\mu\nu}$ and $B_{\mu\nu}$ are those along the spatial directions $i,j=1,2,3,4,5$, i.e., $E_{ij}$ and $B_{ij}$.
Further on, using the $SO(5)$ transformations we can always chose the spatial part of the wave vector $k_\mu$ to be aligned along the fifth direction, thus $k_\mu=(k_0,k_5)$. Then, from \eqref{kf=0} it follows that the only non-zero components of  $B_{ij}$ and  $E_{ij}$ are in the directions orthogonal to $k_5$, i.e. $B_{ab}$ and $E_{ab}$ with $a,b=(1,2,3,4)$. In view of \eqref{k2=0}, which in the frame under consideration is equivalent to $k_0=\pm k_5$, from \eqref{kf=0} we have
\be\label{E=B*}
E_{ab}=\pm\frac 12 \varepsilon_{abcd}B^{cd}\,.
\ee
This solves the null-field condition \eqref{F2=0} and, upon substitution into \eqref{NFsd} produces the non-linear relation
between the components of $B_{ab}$ which solves \eqref{sprelation}.

\section{Exact plane wave and null field solutions in curved space-times}
\label{sec:planewaves}

In this section we will study solutions to the $6D$ Einstein-ModMax system which describe the simultaneous presence and propagation of gravitational and 2-form gauge field waves. We will start with the consideration of pp-waves, interpreted as gravitational waves with planar wave-fronts propagating through Minkowski space-time. Then, we will study  plane waves on top of AdS backgrounds, which correspond to Siklos space-times. Finally, we will explore gauge field waves on top of Robinson-Trautman space-times, which are generically associated with radiative space-times.

\subsection{PP-waves}\label{flatPP}

We look for wave solutions to the Einstein-ModMax equations of motion \eqref{EinsteinMM} with $\Lambda=0$. Let us consider a pp-wave metric\footnote{More information about pp-wave metrics can be found in the introduction of \cite{Roche:2022bcz}.}
\begin{equation}\label{mPP}
    \mathrm{d} s^2= 2\mathrm{d} u\mathrm{d}v+ W(u, x^a)\, \mathrm{d}u^2 + \mathrm{d}x^a \mathrm{d}x^a \, ,
\end{equation}
where\footnote{Note that $\varepsilon_{vu1234}=1$.}  $u=-\frac 1{\sqrt{2}}(x^0-x^5)$, $v=\frac 1{\sqrt{2}}(x^0+x^5)$ (not to be confused with the PST vector field $v_\mu$), and the index $a=1,...,4$ is associated with the flat 4-dimensional Euclidean subspace.
The only nonzero component of the Ricci tensor is
\begin{equation}
    R_{uu} =- \frac{1}{2} \partial_a\partial_a W \,,
\label{Ruu}
\end{equation}
where $\partial_a\partial_a=\partial_a\partial_b \delta^{ab}$ is the Laplacian of the  4-dimensional flat Euclidean space. Let us now make the following ansatz for the 3-form field strength:
\be\label{Fpp1}
F_3=M_{ab}(u)\,\mathrm{d}x^a\wedge \mathrm{d}x^b\wedge \mathrm{d}u\,,
\ee
where $M_{ab}(u)$ is a $u$-dependent $4\times 4$ antisymmetric matrix. Note that $F^2=0$ for this choice.  This is a particular case of null fields, which must satisfy \eqref{sprelation} and \eqref{NFsd} to be a solution of the $6D$ ModMax theory. If we take the auxiliary timelike vector field $v_\mu$ of the PST formalism as
\be\label{vchoice}
v_\mu=\frac 1{\sqrt{1+\frac W2}} \delta_\mu^0\,,
\ee
then the electric field $E_{\mu\nu}$ and magnetic field $B_{\mu\nu}$  have non-zero components only along \hbox{$x^a$-directions}
\begin{align}\label{Enew}
E_2&=\frac{1}{\sqrt{2+ W}}\, M_{ab} \,\mathrm{d}x^a\wedge \mathrm{d}x^b\,, \\
    B_2&=-E_2^*=-\frac 1{2\sqrt{2+ W}} M^{ab}\varepsilon_{abcd}\mathrm{d}x^c\wedge \mathrm{d}x^d=-\frac{1}{\sqrt{2+ W}}\,M^*_{ab}\,\mathrm{d}x^a\wedge \mathrm{d}x^b\,,\label{Bnew}
\end{align}
where the Hodge duality is taken in the four-dimensional Euclidean subspace. This duality relation between $E$ and $B$ is similar to eq. \eqref{E=B*} for plane waves. Substituting the components of \eqref{Enew} and \eqref{Bnew} into the non-linear self-duality
condition \eqref{NFsd} we find that the matrix $M_{ab}(u)$ is constrained as follows
\be\label{MM*}
M_{ab}=-2\cosh\gamma\,\left( M^*_{ab} +\frac{2}{M^*_{cd}M^{*cd}} (M^{*3})_{ab}\right)\,.
\ee
By appropriate $O(4)$ transformations, any antisymmetric $4\times 4$ matrix $M_{ab}$ can be brought to a form in which its only non-zero components are $M^0_{12}=-M^0_{21}$ and $M^0_{34}=-M^0_{43}$. Using this result, we may obtain the general solution of equation \eqref{MM*} in the following form
\be\label{gMab}
M_{ab}=L_a{}^cL_b{}^dM^0_{cd}=2\left(L_{[a}{}^1L_{b]}{}^{2} -e^{\pm\gamma}L_{[a}{}^3L_{b]}{}^{4}\right)\lambda(u)\,,
\ee
where $\lambda(u)=M^0_{12}=-e^{\mp\gamma}M^0_{34}$ is an arbitrary function of $u$ and $L_a{}^{b}(u)$ is a $u$-dependent generic $O(4)$ matrix, i.e. $L_a{}^{c}L_b{}^{d}\delta_{cd}=\delta_{ab}$ $(LL^T=I)$.  Hence,
\be\label{MM2}
M_{ab}M^{ab}=2\lambda^2(u)(1+e^{\pm2\gamma})\,
\ee
and
\begin{equation} \label{Fsquare}
    \frac{2 \pi G_N}{ \cosh\gamma}F_{\mu \rho \sigma} F_{\nu}{}^{\rho \sigma}=\frac{8 \pi G_N\,M_{ab}M^{ab}}{\cosh \gamma \, } \delta_{\mu}^u \delta_{\nu}^u={32 \pi G_N\,e^{\pm\gamma}\lambda^2} \delta_{\mu}^u \delta_{\nu}^u\,,
\end{equation}
Having obtained the solution of the non-linear self-duality condition for the chiral 2-form, we can now look at the corresponding Einstein equations \eqref{EinsteinMM} with $\Lambda=0$. As it turns out, the only component of these equations that is not automatically satisfied is the $uu$-component. Using \eqref{Ruu} and  \eqref{MM2}, we obtain the following necessary and sufficient conditions for the Einstein equations to hold:
\begin{equation}
     \partial_a\partial_a W= -16 \pi G_N \frac{M_{ab}M^{ab}}{\cosh \gamma}=-64  \pi G_N e^{\pm\gamma}\lambda^2(u)\, .
\label{eq:laplaH}
\end{equation}
The solution of \eqref{eq:laplaH} is 
\be\label{H=}
W=C(u) (x^a x^a)^{-1}-8 \pi G_N \lambda^2 e^{\pm\gamma} x^a x^a+c^a(u)x^a+c(u)
\ee
for arbitrary functions $C(u)$, $c_a(u)$ and $c(u)$ of $u$. Observe that $c^a(u)$ and $c(u)$ can be gauged away by means of coordinate transformations.\footnote{To this effect one should perform the shifts $v \rightarrow v-\frac{2c(u)+c^a c^a}{4}-\frac{c_a(u)}{2} x^a$ and $x^a \rightarrow x^a+\frac{1}{2} \int c^a(u)  \mathrm{d}u $.}  When $\gamma=0$,  $L_{a}{}^b=\delta_a{}^b$  and $\lambda$ is a constant, the above solution (for $C(u)=0$) reduces to a maximally supersymmetric solution of $6D$ $\mathcal N=(1,0)$ and $\mathcal N=(2,0)$ supergravities \cite{Meessen:2001vx}.

\subsection{AdS plane-waves}

We will now generalize the result of the previous subsection to  AdS-wave solutions of the Einstein-ModMax system \eqref{EinsteinMM} with the cosmological constant set to $\Lambda= -\frac{10}{\ell^2}$. For that, we will consider the following Siklos space-times \cite{Siklos}, which are interpreted as gravitational waves propagating in the AdS space-time \cite{Gibbons:1986uv,Podolsky:1997ik}:
\begin{equation}
\label{eq:metsiklos}
    \mathrm{d}s^2=\frac{\ell^2}{z^2} \left (2 \mathrm{d}u \mathrm{d}v+\mathrm{d}x^a \mathrm{d}x^a \right) + W(u,x^a) \mathrm{d}u^2\,\,
\end{equation}
where $\ell$ stands for the AdS radius and $x^a=\{x^i,z\}$ $(i=1,2,3)$. The Ricci tensor of \eqref{eq:metsiklos} is
\begin{equation}\label{RTads}
   R_{\mu \nu}=-\frac{5}{\ell^2} g_{\mu \nu} +\frac{1}{2 \ell^2}\left ( 6 W-z^2\partial_a \partial_a W \right) \delta_\mu^u \delta_\nu^u\,.
\end{equation}
Observe that these space-times are Einstein if and only if
\begin{equation}
     6 W-z^2 \partial_a \partial_a W=0\,.
    \label{eq:solsiklos}
\end{equation}
A solution to these equations is given by \cite{Gurses:2014soa}:
\begin{equation}
\label{eq:vsol}
    W(u,x^a)=\sqrt{z} \left (c_1(u) I_{5/2}(z \xi)+c_2(u) K_{5/2}(z \xi) \right) \sin(\vec{\xi} \cdot \vec{x}+c_3(u))\,,
\end{equation}
where $\vec{\xi}$ is a constant vector, $\xi=\vert \vec{\xi} \vert$, $I$ and $K$ stand for the modified Bessel functions and $c_i(u)$ are arbitrary functions of the coordinate $u$. Since \eqref{eq:solsiklos} is linear, the sum or an integral of its solutions with different $\vec{\xi}$ will also be a solution of \eqref{eq:solsiklos}. For $\vec{\xi}=0$, \eqref{eq:vsol} simplifies to $W(u,z)=c(u)z^3$ \cite{Chamblin:1999cj} and the metric describes six dimensional Kaigorodov's space-time \cite{Cvetic:1998jf}.

For the gauge field strength, we take the same ansatz as in \eqref{Fpp1}. Then, it can be readily checked that $M_{ab}(u)$ as in \eqref{gMab} satisfies the ModMax non-linear self-duality equation \eqref{NFsd} in the background \eqref{eq:metsiklos}. Then eq. \eqref{Fsquare} holds again, so the Einstein equations \eqref{RTads} are satisfied
 if and only if
\begin{equation}
\label{eq:edosiklos}
    6 W-z^2 \partial_a \partial_a W =\frac{64 \pi G_N e^{\pm \gamma} \lambda^2 z^4}{\ell^2}\,.
\end{equation}
The solutions to these equations are given by:
\begin{equation}
\label{eq:Vsiklossol}  W(u,x^a)=W^{\mathrm{hom}}(u,x^a)+\frac{32 \pi G_N e^{\pm \gamma} \lambda^2 z^4}{3 \ell^2}\,,
\end{equation}
where $W^{\mathrm{hom}}(u,x^a)$ is the general solution to the homogeneous part of \eqref{eq:edosiklos} given by the linear combination of the solutions \eqref{eq:vsol} with different $\vec{\xi}$.

We have thus derived  asymptotically  AdS plane-wave solutions of the Einstein-ModMax system. The asymptotically flat plane wave solutions considered in Section \ref{flatPP} are obtained  from the AdS ones in the limit $l\to \infty$. To take this limit one should shift the variable $z$ by $l$, i.e. introduce the new coordinate $x^4=z-l$. Then, in the limit $l\to \infty$ the metric \eqref{eq:metsiklos} reduces to \eqref{mPP}, the Ricci tensor \eqref{RTads} reduces to \eqref{Ruu} and the equation \eqref{eq:edosiklos} for $W$ reduces to \eqref{eq:laplaH}.

\subsection{Robinson-Trautman solutions}

Solutions of Robinson-Trautman space-time type \cite{Robinson:1983ws} have been considered in linear p-form theories  coupled to gravity \cite{Ortaggio:2014gma,Ortaggio:2016zuk} and in presence of non-linear vector-field electrodynamics \cite{Tahamtan:2015bha,Kokoska:2021lrn,Kokoska:2022oem,Tahamtan:2023tci}. We will show that the non-linear $6D$ Einstein-ModMax system with a cosmological constant $\Lambda$ also admits such solutions, which generalize the plane-wave solutions considered in the previous subsections.

$D$-dimensional Robinson-Trautman space-times are those admitting a non-twisting, non-shearing but expanding congruence of null geodesics. The class of Robinson-Trautaman space-times includes very different types of solutions, such as static or accelerating black-hole solutions with a cosmological constant or pure radiation solutions (see e.g. \cite{Stephani:2003tm,Podolsky:2006du,Griffiths:2009dfa,Ortaggio:2014gma}). Here we are interested in a radiative subclass of Robinson-Trautaman space-times which are induced by a chiral 2-form gauge field with the ModMax null field strength $F^2=0$ of the form given in \eqref{Fpp1}, \eqref{MM*} and \eqref{gMab}. The form of the Robinson-Trautman metric is inspired by those considered in the context of $D$-dimensional  conformal non-linear  electrodynamics of a gauge vector field in \cite{Kokoska:2021lrn}, namely
\begin{align}
    \label{eq:rtans}
    \mathrm{d} s^2&=2 \mathrm{d} u \mathrm{d}v+W(u,v,x^a
) \mathrm{d} u^2 +h(u,x^a)^{1/4} v^2 \mathrm{d} x^a \mathrm{d} x^a\,,
\end{align}
where $W(u,v,x^a)$ and $h(u,x^a)$ are (a priori) arbitrary functions of the light-cone coordinates $u$ and $v$, and coordinates $x^a$ of four-dimensional Euclidean subspace. As in \cite{Kokoska:2021lrn}, we will take the four-dimensional family of conformally flat Riemannian metrics $h(u,x^a)^{1/4}  \mathrm{d} x^a \mathrm{d} x^a$ be Einstein  with the scalar curvature equal to $4\mu$. Observe that \eqref{eq:rtans} belongs to the Robinson-Trautman class, since
the vector $k^\mu=\delta^\mu_v$ defines a null congruence of geodesics such that
\begin{equation}\label{RTmetric}
    \partial_{[\mu} k_{\nu]}=0\,, \quad \mathcal{L}_{k} g_{\mu\nu}=\frac{\sigma}{2} g_{\mu \nu}+ \xi_{(\mu} k_{\nu)}\,, \quad \sigma=\frac{4}{v}\,, \quad \xi_\mu=\left (\partial_v W- \frac{2 W}{v}\right) \delta_\mu^u -\sigma \delta_\mu^v\,,
\end{equation}
where the expansion $\sigma$ (defined as in \cite{Frolov:2003en}) is clearly non-vanishing.

For the gauge field strength $F_3$, we will take the same ansatz as for the other wave-like solutions, cf. eq. \eqref{Fpp1}.
As before, one finds that the ModMax self-duality condition \eqref{NFsd} is fulfilled if and only if $M_{ab}$ is given by \eqref{gMab}. Hence, we may concentrate on solving the Einstein equations. To this aim, we note that the only non-zero component of the ModMax energy momentum tensor of \eqref{Fpp1} is $T_{uu}$, i.e.
\begin{equation} \label{fsquare}
    \frac{2 \pi G_N}{ \cosh\gamma}F_{\mu \rho \sigma} F_{\nu}{}^{\rho \sigma}= \frac{8 \pi G_N\,M_{ab}M^{ab}}{\cosh \gamma \, v^4 \sqrt{h}} \delta_{\mu}^u \delta_{\nu}^u=\frac{32 \pi G_N\,e^{\pm\gamma}\lambda^2(u)}{ v^4 \sqrt{h}} \delta_{\mu}^u \delta_{\nu}^u.
\end{equation}
Performing the explicit computations or using the results given in the Appendix of \cite{Kokoska:2021lrn}, we find that the $uv$-component of the Einstein equations reduces to
\begin{equation}
    R_{uv}-\frac{1}{2}\Lambda g_{uv}=\frac{1}{2} \left(-\frac{\partial_u h}{v h}+\frac{4 \partial_v W}{v}+\partial_v^2 W-\Lambda\right)=0\,.
\end{equation}
Consequently,
\begin{equation}
    W=\frac{W_1(u,x^a)}{v^3}+W_2(u,x^a)+v \frac{\partial_u h}{4 h}+\frac{v^2}{10}\Lambda\,,
    \label{WRT}
\end{equation}
where $W_1(u,x^a)$ and $W_2(u,x^a)$ are arbitrary functions of the indicated variables. Plugging this result into the   diagonal components of the Einstein equations along the four-dimensional conformally flat Einstein subspace (i.e. $aa$-components), and performing either the explicit calculation or using the results in \cite{Kokoska:2021lrn} one finds that (no summation is implied):
\begin{equation}
    R_{aa}-\frac{1}{2}\Lambda g_{aa}=h^{1/4} (\mu+3 W_2)=0\,, \quad \rightarrow \quad W_2(u,x^a)=-\frac{\mu}{3}\,,
    \label{W2new}
\end{equation}
so $W_2(u,x^a)$ is constant and controlled by the spatial curvature (recall that it is $4\mu$) of the four-dimensional conformally flat sections. Using this result, one learns that:
\begin{equation}
    R_{ua}=-\frac{\partial_a W_1}{2 v^4}=0\,, \quad \rightarrow \quad W_1=W_1(u)\,.
    \label{W1new}
\end{equation}
Finally, using the latter condition in the remaining components of the Einstein equations we find that
\begin{equation}
    -128 \pi G_N e^{\pm\gamma} \lambda^2(u) h^{1/2}+8 h W_1'+5 \partial_u h W_1=0\,.
\end{equation}
The solution of this equation for $h$ is
\begin{equation}
    h=\frac{1}{4 W_1^{8/5}} \left (f(x^a)+\int \frac{128 e^{\pm \gamma} \pi G_N \lambda^2}{5 W_1^{1/5}} \mathrm{d}u \right)^2\,,
    \label{eqHRT}
\end{equation}
where $f(x^a)$ is an arbitrary function of the transverse coordinates $x^a$.  Consequently,  the form of $W(u,v,x^a)$ given by \eqref{WRT}, \eqref{W2new} and \eqref{W1new}, and $h(u,x^a)$ in \eqref{eqHRT} describe a Robinson-Trautman solution of the $6D$ Einstein-ModMax system.

\section{Static  black and extremal dyonic string solutions}
\label{sec:static}

Let us now look for solutions of the Einstein-ModMax system, eqs. \eqref{MMPSTsd} (or \eqref{eomMk'''}) and \eqref{EinsteinMM} in which we set $\Lambda=0$, describing a black self-dual string. To this aim, as suggested by the form of black string solutions in $D=6$ supergravities \cite{Duff:1996hp}, let us assume the following ansatz for the metric:
\be
\label{eq:metdyostring}
    \mathrm{d}s^2_6=\frac 1{g(r)}(-f(r)\mathrm{d}t^2+\mathrm{d}x^2)+g(r)\left (\frac{\mathrm{d}r^2}{f(r)}+r^2\mathrm{d}\Omega^2_3 \right )\,,
\ee
where
$$d\Omega^2_3=\mathrm{d}\theta^2+\sin^2\theta\, \mathrm{d}\phi^2+ \sin^2\theta \sin^2 \phi \, \mathrm{d} \psi^2$$ is the round-sphere metric. For the 2-form gauge field we take
\begin{equation}
    A_2=2 Q K(r) \mathrm{d}t \wedge \mathrm{d}x+2P \sin^2\theta\cos \phi\, \mathrm{d}\theta \wedge \mathrm{d}\psi\,, \label{2-form}
\end{equation}
and
\be
    \label{eq:formdyostring}
    F_3=dA_2= 2Q\, {}^*\epsilon_3 + 2P \epsilon_3\,,
\ee
where \, $\displaystyle \frac{\mathrm{d}K}{\mathrm{d}r}=\frac{1}{r^3 g(r)^2} \, $ ,
$\epsilon_3=\sin^2\theta\sin \phi \, \mathrm{d} \theta \wedge \mathrm{d}\phi  \wedge \mathrm{d}\psi $ \, is the three-sphere volume form and \,
${}^*\epsilon_3=\frac{1}{r^3g(r)^2}\mathrm{d}t\wedge \mathrm{d}x \wedge \mathrm{d}r$ \, is its $6d$ Hodge dual.
The string worldsheet is parameterized by the $(t, x)$-coordinates, and, as we will show below, $Q$ and $P$ are related to the (equal) electric and magnetic charges of the string respectively.

First, let us analyse under which conditions the field strength \eqref{eq:formdyostring} satisfies the self-duality conditions \eqref{MMPSTsd} (or \eqref{eomMk'''}). To this end it is convenient to gauge-fix the auxiliary field $\partial_\mu a=-\delta^0_\mu$. Then, for the ansatz \eqref{eq:formdyostring}, the electric field $E=\frac{1}{2}E_{\mu \nu}\, \mathrm{d}x^\mu \wedge \mathrm{d}x^\nu$ and the magnetic field   $B=\frac{1}{2}B_{\mu \nu}\, \mathrm{d}x^\mu \wedge \mathrm{d}x^\nu$ are\footnote{We are taking the orientation given by $\varepsilon_{txr\theta\phi\psi}=1$.}
\begin{equation}
\label{EB}
E=\frac{2 Q}{r^3 g(r) \sqrt{f(r) g(r)}} \mathrm{d} x \wedge  \mathrm{d} r\,, \quad B=\frac{2 P}{r^3 g(r) \sqrt{f(r) g(r)}} \mathrm{d} x \wedge  \mathrm{d} r\,,
\end{equation}
and one can easily see that the invariant $p$ (defined in \eqref{ps}) is zero. Hence, since $s\geq 0$, the equation \eqref{MMPSTsd} reduces to
\be\label{MMPSTsdr}
E_{\mu \nu}=e^{\gamma} B_{\mu \nu} +\frac{ \sinh\gamma}{s} (B^3)_{\mu \nu}\,,
\ee
which for \eqref{EB} implies
\begin{equation}
\label{eq:chargesstring}
    P=e^{\gamma} Q\,.
\end{equation}
Consequently, the three-form $G_3$ introduced in \eqref{eomMk'''} is
\begin{equation}
\label{eq:gdualsss}
    G_3=2Q e^{\gamma}\, {}^*\epsilon_3 + 2Q \epsilon_3=F^*_3\,.
\end{equation}
Observe that $ F_{\mu\nu\rho}F^{\mu\nu\rho}=\frac{24}{r^6 (g(r))^3} Q^2 \left (e^{2\gamma}-1 \right )$, so it is nonzero unless $\gamma=0$, in which case $F_3$ becomes Hodge self-dual.

Finally, let us relate the constant $Q$ to the physical (Noether) electric and magnetic charges of the string. As usual, these are defined (in our conventions and in accordance with the gauge field equations \eqref{dhatG}), as follows:
\be\label{QeQm}
\mathcal Q_e=\frac{1}{2} \int_{S^3} \iota_{S^3} {} G_3^*\,,\qquad \mathcal Q_m=\frac{1}{2} \int_{S^3} \iota_{S^3}  F_3.
\ee
where $\iota_{S^3}$ stands for the pullback of the embedding of $S^3$ into the six-dimensional space-time. For the ModMax self-dual black string solution we find that
\begin{equation}\label{physQ}
    \mathcal{Q}:=\mathcal Q_e=\mathcal Q_m= 2\pi^2 Q\,e^{\gamma}\,.
\end{equation}
Replacing $Q$ with the physical charge in \eqref{eq:formdyostring} and \eqref{eq:gdualsss} we get
\be\label{FmatchQ}
F_3 = \frac 1{\pi^2} e^{-\gamma}\, \mathcal Q \, {}^*\epsilon_3 + \frac 1{\pi^2} \mathcal Q \,\epsilon_3=G^*_3\,.
\ee
For solving the Einstein equations we compute the ModMax energy momentum tensor. The result is
\be\label{Tmn1}
T_{\mu\nu}=\frac{e^{-\gamma} \mathcal Q^2}{2\pi^4r^6 g(r)^3}(-g_{ab} ,\, g_{ij})\,,
\ee
where $(a,b)$ stand for the indices $(t,x,r)$ and $(i,j)$ are those of the three-sphere. As a consistency check, note that eq. \eqref{eomMk'''} is identically solved by \eqref{FmatchQ} and \eqref{Tmn1}.

Now let us turn to the Einstein equations \eqref{EinsteinMM}. For $\Lambda=0$, these become
\be\label{EMML0}
R_{\mu\nu}-8\pi G_N T_{\mu\nu}=0.
\ee
The components $(ij)$ of these equations along the three-sphere are satisfied if and only if
\begin{align}
   4g^2(1-f)-2r f' g^2+r^2 f(g')^2-rg(rf'g'+3fg'+rfg'') -\frac{8 e^{-\gamma}\mathcal{Q}^2 G_N}{\pi^3 r^{4}}=0\,,
    \label{eq:gpp}
\end{align}
where \emph{prime} stands for the derivative with respect to the radial coordinate. Solving for $g''(r)$ in \eqref{eq:gpp} and substituting the result into the $xx$ component of Einstein's equations \eqref{EMML0}, we get
\begin{equation}
    rf' + 2f-2 =0\,,
\end{equation}
whose solution is
\begin{equation}
    f=1-\frac{r_0^2}{r^2}\,,
    \label{eq:fsol}
\end{equation}
where $r_0^2$ is a constant which we take to be non-negative in order to obtain black configurations. On the other hand, observe that the Ricci scalar associated with the metric \eqref{eq:metdyostring} has the following form:
\begin{equation}
    R=-\frac{g(-6+6f+6 r f'+r^2 f'')+r f(3 g'+r g'')}{r^2 g^2}\,,
    \label{eq:scalarsss}
\end{equation}
Since the tracelessness of the ModMax energy-momentum tensor in \eqref{EMML0} requires that $R=0$, upon the substitution of \eqref{eq:fsol} into \eqref{eq:scalarsss} we find that $R=0$ if and only if
\begin{equation}
   3 g'+r g''=0\,.
   \label{R0}
\end{equation}
The general solution to this linear ordinary differential equation is
\begin{equation}\label{gsol}
    g=g_\infty+\frac{g_0}{r^2}\,.
\end{equation}
After an appropriate rescaling of the coordinates one can set $g_\infty=1$, which we will further assume. This form of $g(r)$ fixes $\displaystyle K(r)=\frac{1}{2g_0g(r)}$ in \eqref{2-form} modulo a constant term.

Now Einstein's equations \eqref{EMML0}
for the metric \eqref{eq:metdyostring} with $f(r)$ and $g(r)$ given in \eqref{eq:fsol} and \eqref{gsol}, and $T_{\mu\nu}$ given in \eqref{Tmn1}
are all satisfied if
\begin{equation}
r_0^2=-g_0+ \frac{2\,G_N\,e^{-\gamma} \mathcal Q^2}{\pi^3 g_0}\,.
\label{eq:chargesstring2}
\end{equation}
Here $r=r_0$  is the location of the event horizon,
and the functions $f$ and $g$ now become
\begin{equation}\label{fg1}
    f=1+\frac{1}{r^2}\left(g_0-\frac{2\,G_N\,e^{-\gamma} \mathcal Q^2}{ \pi^3g_0}\right)\,, \quad g=1+\frac{g_0}{r^2}\,.
\end{equation}
We will take $g_0 \geq 0$ to avoid curvature singularities in the region $r>0$. Observe that whenever $r_0^2>0$, the solution possesses a regular non-extremal horizon placed at $r=r_0$, whose near-horizon geometry is $\mathrm{Rindler}_2 \times \mathbb{R}\times S^3$, while in the case $r_0=0$
near $r=0$ the geometry is AdS$_3 \times S^3$, exactly as in the case of the corresponding extremal dyonic-string solutions in $D=6$ supergravities.

The computation above shows that, equations \eqref{eq:metdyostring}, \eqref{eq:formdyostring}, \eqref{eq:chargesstring} and \eqref{fg1} describe a solution of the Einstein-ModMax system of equations. From the form of $f(r)$ in \eqref{fg1} we see that this solution is endowed with an event horizon whenever
\begin{equation}
 g_0^2\leq  e^{-\gamma}\,\frac{ 2\,G_N\,\mathcal Q^2  }{\pi^3}\,.
    \label{eq:boundbh}
\end{equation}
For $\gamma=0$ and when the bound is not saturated, this black-string solution coincides with that considered in \cite{Duff:1995yh} which can be oxidized to a solution of the NS-NS sector of string theory (see e.g. \cite{Ma:2021opb}).\footnote{For $\gamma=0$, our charge $\mathcal Q$ is related to the analogous charge $Q_e$ of \cite{Ma:2021opb} as follows $\mathcal Q=\sqrt{\frac{8\pi}{ G_{N}}}Q_e$.} When the bound is saturated, i.e.
\be\label{satur}
g_0^2=e^{-\gamma}\,\frac{ 2\,G_N\,\mathcal Q^2  }{\pi^3} \,,
\ee
we have $f=1$ from \eqref{fg1}, and the horizon shrinks to zero since now $r_0=0$ \eqref{eq:chargesstring2}. This solution describes the ModMax counterpart of the solitonic self-dual string of $D=6$ supergravities preserving half supersymmetry \cite{Duff:1995yh}, the supergravity solution being recovered at $\gamma=0$. Note that for $\gamma >0$ the effect of $e^{-\gamma}$  is to decrease the bound on $g_0^2$, similar to the effect of ModMax on the bound of the mass of Reissner-Nordstr\"om black holes in $D=4$ \cite{Flores-Alfonso:2020nnd,BallonBordo:2020jtw,Flores-Alfonso:2020euz}. We will see below how this affects the ADM mass of the string.

Let us now briefly consider the thermodynamics associated with the ModMax black string configuration. The black-string temperature $T_{\mathrm{BS}}$ is  computed as the periodicity of the Euclidean time that ensures the regularity of the Euclidean geometry. The result is
\begin{equation}
    T_{\mathrm{BS}}=\frac{r_0}{2 \pi(r_0^2+g_0)}=\frac{\pi^2 g_0}{4G_Ne^{-\gamma}\mathcal Q^2}\sqrt{\frac{2G_Ne^{-\gamma}\mathcal Q^2}{\pi^3 g_0}-g_0}\,,
\end{equation}
Regarding the black-string entropy $S_{\mathrm{BS}}$ (per unit length in the $x$-direction of the string), it is equal to the one fourth of the area of the event horizon $A=\int_{S^3} r_0^3 g(r_0)\epsilon_3 $ divided by $G_N$, since there are no higher-curvature corrections involved:
\begin{equation}
    S_{\mathrm{BS}}=\frac{\pi^2}{2{G_N}}r_0^3 \left (1+\frac{g_0}{r_0^2} \right)= \frac{e^{-\gamma}\mathcal Q^2}{\pi g_0}\sqrt{\frac{2G_Ne^{-\gamma}\mathcal Q^2}{\pi^3 g_0}-g_0}\,.
\end{equation}
Finally, the ADM mass density $M_{\mathrm{BS}}$ along the  $x$-direction (i.e., the string tension $T$), computed using the prescription described in \cite{Lu:1993vt}, is
\begin{equation}
    M_{\mathrm{BS}}=\frac{\pi}{2G_N} \left ( g_0+\frac{3 r_0^2}{4}\right)=\frac{\pi}{8G_N}\left(g_0+\frac{6G_Ne^{-\gamma}\mathcal Q^2}{\pi^3g_0}\right)\,.
\end{equation}
Note \emph{en passant} that when $r_0=0$, which is the case of the extremal string solution,
\be\label{Mextreme}
 M^{extr}_{\mathrm{BS}}=\frac{\pi}{2G_N}g_0=e^{-\frac\gamma 2}\frac{2|\mathcal Q|}{\sqrt{8\pi G_N}}.
\ee
This implies that the ratio of the ADM mass to the charge of the extremal (solitonic) ModMax string differs from that of the related supergravity solution by a factor of $e^{-\gamma/2}$.  The effect of the factor $e^{-\gamma/2}$ in this solution resembles the effect of a constant dilaton $e^{\phi_0}$ in a self-dual string solution of a $6D$ supergravity discussed in \cite{Duff:1993ye,Duff:1996cf}.

In order to arrive at the first law of black-hole thermodynamics, one needs to derive the corresponding chemical potentials (for details see e.g. \cite{Ortin:2022uxa}). On the one hand, the `electric' chemical potential  $\Phi_{\rm e}$ associated with $F_3=dA_2$ is
\begin{equation}
\label{eq:cheme}
    \Phi_{\rm e}=\lim_{r \rightarrow +\infty} (A_2)_{tx} - \lim_{r \rightarrow r_0} (A_2)_{tx}=\frac{ Q}{(g_0+r_0^2)}=\frac{\pi g_0}{4G_N\mathcal Q}\,,
\end{equation}
On the other hand, since due to eq.  \eqref{eomMk} we have $F_3=G_3^*$, then locally $G_3^*=\mathrm{d}{A}_2$, and the  `magnetic' chemical potential $\Phi_{\rm m}$ associated with $G_3^*$ is equal to $\Phi_e$.
\if{}
\begin{equation}
    \Phi_{\rm m}= \lim_{r \rightarrow +\infty} (A_2)_{tx} - \lim_{r \rightarrow r_0} (A_2)_{tx}=\Phi_{\rm e}\,.
\end{equation}
\fi
 Then, denoting the common value of the chemical potential by $\Phi$, we find that the following first law of black-string thermodynamics is satisfied
\begin{equation}
\mathrm{d}M_{\mathrm{BS}}=T_{\mathrm{BS}} \mathrm{d}S_{\mathrm{BS}}+ 2\Phi \mathrm{d} \mathcal{Q}\,.
\end{equation}
The factor of two in front of the chemical potential  is due to the equality  between $(\mathcal Q_e, \Phi_{\rm e})$ and $(\mathcal Q_m,\Phi_{\rm m})$ (a consequence of the ModMax self-duality condition).

\subsection{Extremal wavy string}
\label{strw}

Additionally, let us show that there also exist solutions describing extremal dyonic ModMax strings with  waves propagating along their worldsheet similar to those considered in \cite{Garfinkle:1992zj}. We make the following ansatz for the metric
\be
\label{eq:metextwave}
  \mathrm{d}s^2_6=\frac 1{g(r)}(2 \mathrm{d}u \mathrm{d}v+h(u,r) \mathrm{d}u^2)+g(r)\left (\mathrm{d}r^2+r^2\mathrm{d}\Omega^2_3 \right )\,,
\ee
demanding $A_2$ and $F_3$ to be as in \eqref{2-form} and \eqref{eq:formdyostring}. We also take $P$ and $Q$ to be related by the ModMax non-linear self-duality condition \eqref{eq:chargesstring} and to the physical string charge $\mathcal Q$ by \eqref{physQ}.

Computing the Ricci scalar of \eqref{eq:metextwave} and setting it to zero, we obtain the same equation for $g(r)$ as in the static black string case \eqref{R0} which implies
\begin{equation}
     g(r)=1+\frac{g_0}{r^2}\,.
\end{equation}
Now we  observe that, as in the static extremal string case, the angular (three-sphere) components of the Einstein equations \eqref{EinsteinMM} with $\Lambda=0$ are solved if and only if $g_0$ is related to the string charge $\mathcal Q$ as in \eqref{satur}.
Then, all the components of the Einstein equations are solved, with the exception of the $uu$-component. The condition we get for its vanishing is
\begin{equation}
    3 \partial_r h +r \partial_r^2 h=0\,.
\end{equation}
The most general solution to this equation is $h(u,r)=c(u)+\frac{f(u)}{r^2}$, for arbitrary $u$-dependent functions $c$ and $f$. However, the function $c(u)$ can be gauged away by a redefinition of the $v$-coordinate, and we are left with
\begin{equation}
    h(u,r)=\frac{f(u)}{r^2}\,.
\end{equation}
Thus, we have shown that the $6D$ Einstein-ModMax system admits an extremal self-dual string solution with waves propagating along its worldsheet.

\section{Towards rotating strings in $6D$ Einstein-ModMax theory}

\label{sec:rotbh}
Finally, we would like to conclude this study of solutions of the $6D$ Einstein-ModMax system with a preliminary result towards the finding of the first rotating ModMax black-string solutions that would generalize those found in $D=6$ supergravities \cite{Cvetic:1998xh,Cvetic:1996xz,Cvetic:2013cja,Chow:2019win,Pang:2019qwq}. As it turns out, deriving a complete solution of this type proves to be a rather non-trivial problem. A similar issue has been encountered in four dimensions \cite{Kubiznak:2022vft}, where it was shown that a canonical slowly rotating black-hole ansatz does not solve the equations of motion of generic non-linear electrodynamics theories coupled to GR.

So far, we have managed to find an exact solution which we interpret as a \emph{`near horizon'} limit of a rotating self-dual extremal Einstein-ModMax string. As an ansatz for the metric and the three-form field strength of this solution, we take the one similar to that considered for rotating black strings in $6D$ supergravity \cite{Chow:2019win}:
\begin{align}
\label{eq:rotmet}
    \mathrm{d}s^2&=A(r)\left (-\mathfrak{e}_1 \otimes \mathfrak{e}_1+\mathfrak{e}_2 \otimes \mathfrak{e}_2 \right)+\frac{\mathrm{d}r^2}{D(r)}+\frac{r^2}{4 D(r)}\left (\sigma_3^2+ \mathrm{d}\theta^2+\sin^2\theta \mathrm{d}\phi^2 \right)\,,\\
   F_3&=\frac{\mathcal Q\,e^{-\gamma}}{\pi^2}\,\mathrm{d}\left ( K(r) \, \mathrm{d}t \wedge \mathrm{d}x\right)+\frac{\mathcal Q}{8\pi^2}\sin \theta \, \mathrm{d}\theta \wedge \mathrm{d}\phi \wedge \mathrm{d} \psi+\mathrm{d}\left ( W(r) \, (\mathrm{d}t - \mathrm{d}x) \wedge \sigma_3\right) \,,\\
    \mathfrak{e}_1&=\mathrm{d}t+\omega(r) \sigma_3\,, \quad \mathfrak{e}_2=\mathrm{d}x+\omega(r) \sigma_3\,, \quad \sigma_3=\mathrm{d}\psi-\cos \theta \mathrm{d}\phi\,.
\end{align}
where $\mathcal Q$ is associated with the string charge as in \eqref{QeQm} and \eqref{physQ}. For this solution, we found convenient to perform a change of the angular coordinates with respect to the static string case, to match directly with similar results in the literature \cite{Chow:2019win}. Namely, here $S^3$ is parameterized as an $S^1$ fibre over $S^2$.

We find that the Einstein equations \eqref{EinsteinMM} (with $\Lambda=0$) and the ModMax equations \eqref{eomMk'''} are satisfied if the functions $A(r)$, $D(r)$, $K(r)$, $W(r)$ and $\omega(r)$ have the following form:
\begin{equation}
\begin{split}
    A(r)=D(r)&=\frac{2\pi^2 e^{\gamma/2}r^2}{\sqrt{8 \pi G_N}\mathcal Q}\,, \qquad K(r)=\frac{\pi^3 e^{\gamma}r^2}{ 4G_N \mathcal Q^2}\,,  \\ W(r)&=\frac{\pi J}{8 \mathcal Q}e^{\gamma/2}\,, \qquad \omega(r)=\frac{J{G_N} e^{-\gamma/2}}{2r^2}\,,
    \end{split}
    \label{eq:nhmodmaxrot1}
\end{equation}
where $J$ is an arbitrary constant of dimension $length^{-1}$ which we will relate below to string angular momentum. Observe that, in the limit $J \rightarrow 0$, one recovers the near-horizon geometry of the static extremal dyonic string of Section \ref{sec:static}.

For the functions \eqref{eq:nhmodmaxrot1}, it turns out that the following shift of the angular \linebreak \hbox{coordinate $\psi$}
\begin{equation}
  \mathrm{d}\psi=\mathrm{d}\hat\psi  + \frac{\pi^3 e^{\gamma/2} J}{ \mathcal Q^2}(\mathrm{d}t-\mathrm{d}x)
\end{equation}
brings the metric \eqref{eq:rotmet} to the form
\begin{equation}
\label{eq:nhmet1}
    \begin{split}
    \mathrm{d}s^2&=\frac{4\pi^2 \,e^{\gamma/2}}{\sqrt{8 \pi G_N}\,\mathcal Q}\,r^2\, \mathrm{d}u \mathrm{d}v-\frac{\pi^2 \sqrt{8\pi G_N} J^2\,e^{\gamma/2}}{4\mathcal Q^3} \mathrm{d}u^2+\frac{\sqrt{8 \pi G_N}\,\mathcal Q\,e^{-\gamma/2} } {2\pi^2 }\,\frac{\mathrm{d}r^2}{r^2}\\&
    +\frac{\sqrt{8 \pi G_N}\,\mathcal Q e^{-\gamma/2}}{8\pi^2} \left ((\mathrm{d}\hat\psi- \cos \theta \mathrm{d}\phi)^2+ \mathrm{d}\theta^2+\sin^2\theta \mathrm{d}\phi^2\right)\,,
    \end{split}
\end{equation}
where, as in the case of wave solutions, $u=-\frac 1{\sqrt{2}}(t-x)$ and $v=\frac 1{\sqrt{2}}(t+x)$. One can see that the metric \eqref{eq:nhmet1} corresponds to an $\mathrm{AdS}_3 \times S^3$ geometry, characteristic of extremal near-horizon geometries of six-dimensional strings, while the AdS part of the metric actually corresponds to the rotating BTZ black hole \cite{Banados:1992wn}. It can be checked that for $\gamma = 0$, the metric \eqref{eq:nhmet1} describes the near-horizon geometry of a supersymmetric  rotating extremal self-dual string with equal electric and magnetic charge considered in \cite{Cvetic:1998xh, Chow:2019win}.\footnote{Strictly speaking, the \emph{would-be} horizon in that solution  turns out to be singular, since it corresponds to a curvature singularity. Nevertheless, we may understand this solution as the extremal limit of a proper black solution with a regular horizon. In this sense, one may regard the geometry around $r=0$ as near-horizon.}  From this comparison, we infer that in our solution $J$ is associated with the angular momentum per unit length of the string. To guarantee that this is the proper physical interpretation of this constant, one should find the fully fledged rotating solution and analyse its asymptotic behaviour, a feature which we miss in the near-horizon solution.

Consequently, it is natural to interpret the metric \eqref{eq:nhmet1}  as the near-horizon geometry of a rotating string solution in the $6D$ Einstein-ModMax system which is yet to be found.

\section{Conclusions and Outlook}
In this paper we have presented the first examples of solutions of $6D$ ModMax theory coupled to gravity.
We have concentrated on the study of waves and self-dual black string solutions. On the one hand, we analysed pp-wave, Siklos and Robinson-Trautman spacetimes, determining how the ModMax parameter $\gamma$ affects solutions with $\gamma=0$. On the other hand, we studied black-string configurations and how they differ from similar six-dimensional supergravity solutions. In particular, we found a static self-dual black string solution and showed that it satisfies the first law of thermodynamics in which the  dependence of the ADM mass, temperature and entropy of the black string on the string charge is screened by a factor of $e^{-\gamma/2}$ of the positive ModMax parameter $\gamma$. Interestingly enough, the effect of the factor $e^{-\gamma/2}$ in this solution resembles the effect of a constant dilaton $e^{\phi_0}$ in a self-dual string solution of a $6D$ supergravity discussed in \cite{Duff:1993ye,Duff:1996cf}. It would be of interest to elaborate on this analogy elsewhere.

There are several directions one may take for further study of this theory. For instance, a non-trivial problem is to find a fully fledged rotating extremal string solution in the Einstein-ModMax system, whose near horizon limit was obtained in \hbox{Section \ref{sec:rotbh}}.  Another interesting problem is the full classification of solutions of the Einstein-ModMax system in the class of Robinson-Trautman spacetimes with an aligned chiral three-form field strength, generalizing those obtained for $\gamma=0$ case in \cite{Ortaggio:2014gma}. In yet other perspective, there is a solution generating technique developed by Garfinkle and Vachaspati \cite{Garfinkle:1990jq, Garfinkle:1992zj} within the context of supergravities which allows one to add waves to an existing solution possessing a null Killing vector without changing other physical fields. The string solution with propagating waves that we found in Subsection \ref{strw} suggests that this method might also be generalized to $6D$ ModMax.

Another venue one can follow is to examine the dimensional reduction of the found solutions and analyse the corresponding solutions one gets in the four-dimensional ModMax theory or its further reduction to three space-time dimensions, and compare these results with those known in the literature (see e.g. \cite{Flores-Alfonso:2020euz,BallonBordo:2020jtw,Flores-Alfonso:2020nnd,Bokulic:2021dtz,Barrientos:2022bzm,Bakhtiarizadeh:2023mhk}).

One more interesting problem is the supersymmetrization of the $6D$ ModMax chiral 2-form theory and its coupling to $N=(1,0)$, $D=6$ supergravity. In four space-time dimensions, an $N=1$ superconformal generalization of ModMax electrodynamics was constructed and coupled to $N=1$, $D=4$ supergravity \cite{Kuzenko:2021cvx,Bandos:2021rqy} with the use of the $N=1$ superfield formalism, crucial for the accomplishment of this construction. An $N=2$ supersymmetric extension of the $4D$ ModMax has not been completely successful yet,
the main technical difficulty being the fact that the full ${ N}=2$ superspace measure is dimensionless and so must be the superspace Lagrangian density as well, thus causing a problem with having  ${ N}=2$ superconformal invariance.
In $D=6$ the simplest supersymmetry is $N=(1,0)$ which is related to $N=2$, $D=4$ supersymmetry by dimensional reduction. We therefore expect that the problem of the supersymmetrization of the $6D$ ModMax and its coupling to supergravity will be rather challenging, but exciting, also in view of the fact that the superfield description of $N=(1,0)$, $D=6$ supersymmetric theories is much more involved than of the $N=1$, $D=4$ ones (see \cite{Kozyrev:2022dri} for a Lagrangian description of a linear chiral 2-form multiplet in an $N=(1,0)$, $D=6$ harmonic superspace \cite{Howe:1985ar,Sokatchev:1988aa}). As a starting point one may study the possibility of coupling the ModMax (anti)-chiral 2-form to another  chiral 2-form field and a dilaton, the fields which, together with the graviton, constitute the bosonic part of the $N=(1,0)$ supergravity coupled to a linear anti-chiral 2-form multiplet (see Appendix). As an alternative approach to the supersymmetrization problem, one may try to apply $T\bar T$-like deformation techniques \cite{Ferko:2023ruw,Ferko:2024zth} to the component action of \cite{Dall'Agata:1997db} for the linear chiral 2-form supermultiplet, facing however the problem of on-shell closure of the supersymmetry transformations.

In another vein, it would also be interesting to explore solutions of the $6D$ ModMax system including higher-order curvature corrections. To this aim, one could minimally couple ModMax to higher-curvature gravities, such as (Generalized) Quasitopological Gravities \cite{Oliva:2010eb,Myers:2010ru,Hennigar:2017ego, Bueno:2022res,Moreno:2023rfl} or their upgraded cosmological versions \cite{Moreno:2023arp}. One can also examine the inclusion of non-minimal gravitational couplings of ModMax following the strategy of \cite{Cano:2020qhy,Cano:2021tfs,Cano:2022ord}.

\section*{Acknowledgements}
The authors are grateful to Igor Bandos, Sergei Kuzenko and Kurt Lechner for useful discussions and comments. Work of \'A. J. M. has been supported by a postdoctoral grant from the Istituto Nazionale di Fisica Nucleare, Bando 23590. D.S. acknowledges financial support from an ICTP grant, by the Australian Research Council, project No. DP230101629, by the Spanish AEI MCIN and FEDER (ERDF EU) under grant PID2021-125700NB-C21 and by the Basque Government Grant IT1628-22. N.S.D. wishes to thank INFN Padova for financial support and warm hospitality during the last phase of this project. D.S. acknowledges the very kind hospitality extended to him at the Feza Gursey Center for Physics and Mathematics (Istanbul), where this project was initiated between 25 Oct and 8 Nov 2023.

\appendix
\section{The relation between two formulations of \hbox{$N=(1,0)$,} $D=6$ supergravity coupled to a single tensor multiplet}
The $N=(1,0)$, $D=6$ supergravity under consideration has two supermultiplets, the gravitational multiplet which contains the graviton $g_{\mu\nu}(x)$,  a chiral 2-form field $A_{\mu\nu}^0$ and a gravitino $\psi^\alpha_\mu(x)$, and the tensor supermultiplet which contains an anti-chiral 2-form field $A^1_{\mu\nu}$, a scalar (dilaton) field $\phi(x)$, and a spinorial field $\chi^\alpha(x)$ (the index $\alpha$ labels a symplectic Majorana-Weyl spinor representation of the $6D$ Lorentz group). There are two Lagrangian formulations of this theory. In the first one  the chiral and anti-chiral 2-form field are combined into a single non-chiral field $\tilde A_{\mu\nu}$ for which the action has the conventional form \cite{Marcus:1982yu}. In the second formulation the chiral and anti-chiral field are treated as a doublet $A^I_{\mu\nu}=(A^0_{\mu\nu},A^1_{\mu\nu})$, and the action is constructed with the use of the PST technique \cite{Dall'Agata:1997db}. We would like to elucidate the relation between these two formulations.

The bosonic part of the action of \cite{Marcus:1982yu} in our normalization of fields (and setting the gravitational constant $8\pi G_N=1$) is
\be\label{sgactionH}
\mathcal S=\int d^6x \sqrt{-g}\left(\frac 12 R-\frac12\partial_\mu\phi \partial^\mu\phi-\frac1{24}e^{-2\phi}H_{\mu\nu\rho}H^{\mu\nu\rho}\right)\,,
\ee
where $H_{\mu\nu\rho}=3\partial_{[\mu}\tilde A_{\nu\rho]}$.

On the other hand, in the PST formulation the bosonic part of the $N=(1,0)$, $D=6$ supergravity action containing one chiral and one anti-chiral 2-form has the following form \cite{Dall'Agata:1997db} in our conventions and notation
\be\label{sgaction}
\mathcal S=\int d^6x \sqrt{-g}\left(\frac 12 R-\frac12\partial_\mu\phi \partial^\mu\phi+\frac14\left(\eta_{IJ}E^I_{\mu\nu}B^{J\,\mu\nu}-G_{IJ}(\phi)B^I_{\mu\nu}B^{J\,\mu\nu}\right)\right)\,,
\ee
where  $B^I_{\mu\nu}$ and $E^I_{\mu\nu}$ are as defined in \eqref{Bp} and $\eqref{E}$ for each of the two tensor field strengths $F^I_3=dA^I_3$ $(I=0,1)$,
\be\label{etaIJ}
\eta_{IJ}=\left(
\begin{matrix}
1 & \,0\\
0 & \, -1
\end{matrix}
\right)\,,
\ee
and
\be\label{G}
G_{IJ}(\phi)=\left(
\begin{matrix}
\cosh(2\phi)&\,\, \sinh(2\phi)\\
\sinh(2\phi)& \,\, \cosh(2\phi)
\end{matrix}
\right)\,.
\ee
An $SO(1,1)$ group element matrix $L_J{}^{\underline I}=( L_J, L_J{}^{1})$ has the following form
\be\label{Lmatrix}
L_I{}^{\underline J}=
\left(
\begin{matrix}
\cosh\phi &\, \sinh \phi\\
\sinh \phi &\, \cosh \phi
\end{matrix}
\right)
\ee
So $\eta_{IJ}=L_I{}^{\underline K}\eta_{\underline{KL}}L_J{}^{\underline L}$,
$G_{IJ}=L_I{}^{\underline K}\delta_{\underline{KL}}L_J{}^{\underline L}$ and
\be\label{GG-1}
(G_{IJ})^{-1}=\eta^{IK}G_{KL}\eta^{LJ}.
\ee
The `twisted' self-duality relations which one gets by varying the above action with respect to $A^I_{\mu\nu}$ and solving the corresponding equations of motion have the following form
\be\label{twistedsd}
E^I_{\mu\nu}=\eta^{IK}G_{KJ}(\phi)B^J_{\mu\nu}\,,
\ee
which are equivalent to
\be\label{twistedsd1}
F^I_{\mu\nu\rho}=\eta^{IK}G_{KJ}(\phi)F^{*J}_{\mu\nu\rho}\,, \qquad F^{*I}_{\mu\nu\rho}=\eta^{IK}G_{KJ}(\phi)F^{J}_{\mu\nu\rho}\,.
\ee
They are consistent with the Hodge duality operation because of the property \eqref{GG-1} of $G_{IJ}$. Note that for $\phi=0$ these equations disentangle and reduce to the self-duality of $F_3^0$ and the anti-self-duality of $F_3^1$
\be\label{phi=0}
F^0_{\mu\nu\rho}=F^{*0}_{\mu\nu\rho}\,,\qquad F^1_{\mu\nu\rho}=-F^{*1}_{\mu\nu\rho}\,.
\ee
Let us now show how the self-duality relations \eqref{twistedsd1} can be combined to produce the equation of motion for a non-chiral 2-form field strength which is obtained from the action \eqref{sgactionH}. To this end we rewrite \eqref{twistedsd1} for each value of $I=(0,1)$
\bea\label{twistedsd2}
F^{*0}_{\mu\nu\rho}&=&\cosh(2\phi) F^0_{\mu\nu\rho}+\sinh(2\phi)F^1_{\mu\nu\rho}\nonumber\\
F^{*1}_{\mu\nu\rho}&=&-\sinh(2\phi) F^0_{\mu\nu\rho}-\cosh(2\phi)F^1_{\mu\nu\rho}\,.
\eea
Taking the sum of the above relations we have
\be\label{F+F}
e^{-2\phi}(F^0_{\mu\nu\rho}-F^1_{\mu\nu\rho})=F^{*0}_{\mu\nu\rho}+F^{*1}_{\mu\nu\rho}\,.
\ee
Acting now on the both sides with the covariant derivative $D_\mu$ we get
\be\label{DH}
\partial_\mu\left(\sqrt{-g}\,e^{-2\phi}(F^{0\,\mu\nu\rho}-F^{1\,\mu\nu\rho})\right)=0\,.
\ee
This is exactly the equation of motion  for the non-chiral gauge field which one obtains from the action \eqref{sgactionH} if we set
\be\label{H=F-F}
H_{\mu\nu\rho}=F^0_{\mu\nu\rho}-F^1_{\mu\nu\rho}, \qquad \tilde A_{\mu\nu}=A^0_{\mu\nu}-A^1_{\mu\nu}.
\ee
The relation \eqref{F+F} is just the solution of the equation \eqref{DH} in terms of the Hodge dual of the second linear combination (i.e. the sum) of $F^0_3$ and $F^1_3$.

Let us now derive the energy-momentum tensor of the fields $F^I_{\mu\nu\rho}$ by varying the action \eqref{sgaction} with respect to the metric $g^{\mu\nu}$. The result is
\be\label{EMPST1}
T_{\mu\nu}=\frac 14 g_{\mu\nu}G_{IJ}\,B^I_{\mu\nu}B^{J\mu\nu}+\frac 12 v_\mu v_\nu G_{IJ}\,B^I_{\mu\nu}B^{J\mu\nu} +\frac 14 G_{IJ}(B^IB^J)_{\mu\nu}- 2 v_{(\mu}p_{\nu)},
\ee
where now
$$
p^\mu=-\frac 18\varepsilon^{\mu\nu\rho\sigma\lambda\kappa}\left(B^I_{\nu\rho}B^J_{\sigma\lambda}v_\kappa\right) \eta_{IJ}\,.
$$
On the mass-shell \eqref{twistedsd} the energy momentum tensor reduces to that of $H_{\mu\nu\rho}$
\bea\label{TI}
T_{\mu\nu}|_{on-shell}&=&\frac 14F^I_{\rho\lambda(\mu}F_{\nu)}^{*J\,\,\rho\lambda}\eta_{IJ}\nonumber\\
&=&\frac 14e^{-2\phi}\left(H_{\mu\rho\lambda}H_\nu{}^{\rho\lambda}-\frac 16g_{\mu\nu} H^2\right),
\eea
where to derive the expression on the right hand side one should use the relation \eqref{F+F} and the definition \eqref{H=F-F} of $H_{\mu\nu\rho}$.

One can also directly verify that the dilaton equations of motion match as well.
Indeed, the dilaton field equation which follows from \eqref{sgactionH} is
\be\label{dem}
D_\mu D^\mu\phi=-\frac {1}{12}e^{-2\phi}H_{\mu\nu\rho}H^{\mu\nu\rho}\,,
\ee
while that produced by \eqref{sgaction} is
\be\label{demPST}
D_\mu D^\mu \phi=\frac 14 \frac {dG_{IJ}(\phi)}{d\phi}B^{I}_{\mu\nu}B^{J^{\mu\nu}}=\frac 14 e^{2\phi}(B_{\mu\nu}^0+B_{\mu\nu}^1)^2-\frac 14 e^{-2\phi}(B_{\mu\nu}^0-B_{\mu\nu}^1)^2\,.
\ee
The latter, using eq. \eqref{F+F} projected along $v^\mu$ and the definition of \eqref{H=F-F}, can be rewritten as
\bea
D_\mu D^\mu\phi=&&\frac 14 e^{-2\phi}\left[(E_{\mu\nu}^0-E_{\mu\nu}^1)^2-(B_{\mu\nu}^0-B_{\mu\nu}^1)^2\right]\\
=&&\frac 14 e^{-2\phi}(H_{\mu\nu\rho}v^{\rho}H^{\mu\nu\lambda}v_{\lambda}-H^*_{\mu\nu\rho}v^{\rho}H^{*\mu\nu\lambda}v_{\lambda})\equiv -\frac {1}{12}e^{-2\phi}H_{\mu\nu\rho}H^{\mu\nu\rho},\nonumber
\eea
where to arrive at the final equality we used the identity \eqref{id} applied to $H_{\mu\nu\rho}$.

We have thus shown that the bosonic sectors of the supergravity theories described by the action \eqref{sgactionH} and \eqref{sgaction} are classically equivalent. The addition of the fermionic sectors to these formulations would produce only some technical complications for their comparison without changing the result.


\providecommand{\href}[2]{#2}\begingroup\raggedright\endgroup

\end{document}